# Plasma properties from the multi-wavelength analysis of the November 1st 2003 CME/shock event

**Carlo Benna[1], Salvatore Mancuso[1], Silvio Giordano[1], Lorenzo Gioannini[2]**

[1] *Istituto Nazionale di Astrofisica (INAF) - Osservatorio Astrofisico di Torino (OATo), Italy*

[2] *Dipartimento di Fisica Generale, Università degli Studi di Torino, Italy*

## Abstract

The analysis of the spectral properties and dynamic evolution of a CME/shock event observed on November 1st 2003 in white-light by the LASCO coronagraph and in the ultraviolet by the UVCS instrument operating aboard SOHO, has been performed to compute the properties of some important plasma parameters in the middle corona below about 2 $R_\odot$. Simultaneous observations obtained with the MLSO/Mk4 white-light coronagraph, providing both the early evolution of the CME expansion in the corona and the pre-shock electron density profile along the CME front, were also used to study this event. By combining the above information with the analysis of the metric type II radio emission detected by ground-based radio spectrographs, we finally derive estimates of the values of the local Alfvén speed and magnetic field strength in the solar corona.

## Introduction

Multi-wavelength observations of coronal mass ejections (CMEs) can be succesfully used to investigate the main physical plasma parameters of the solar corona. For the most energetic CME events, a forward fast-mode magnetohydrodynamic (MHD) shock is expected to be produced ahead of the front. Coronal shocks are thought to be driven by coronal mass ejections and/or by flare ejecta. Alternatively, their origin can be attributed to flare-ignited blast waves. However, even after several decades, the origin of coronal shock waves is not completely understood as yet (see, e.g., [1] and references therein). Generally, CME-driven shocks are too faint to be detected by white-light coronagraphs in the visible range, but they can be easily identified in radio dynamic spectra as type II radio bursts, that is, narrow bands of enhanced radio emission generated near the local electron plasma frequency $f_{pe}$ and/or its harmonics. However, due the lack of information about the location of the type II emission in the corona, it is important to integrate the data with ancillary observations from instruments having also spatial resolution. In the past decade, spectroscopic observations from the SOHO's UltraViolet Coronagraph Spectrometer (UVCS) [2] have been effectively used for coronal shock detection [3-8]. UV emission lines can be very important sources of diagnostic information about the physical properties of the solar corona. In fact, line intensities and profiles critically depend on several parameters, such as the electron density, the electron temperature, the kinetic temperature of the emitting ions and the outflow velocity. In the present paper, we analyze multi-wavelength observations of a CME/shock event observed on November 1$^{st}$ 2003 at 1.7 $R_\odot$ beyond the southwest limb of the Sun. The main goal of this work is to retrieve information on the local Alfvén Mach number and magnetic field strength of the pre-shock plasma in the middle corona.

**Event description and data coverage**

A candidate CME/shock event observed on 2003 November 1$^{st}$ included in the online CDAW LASCO CME catalog (http://cdaw.gsfc.nasa.gov/CME_list/index.html) [9] was selected by using a UVCS catalog of CMEs [10] composed by over 1000 events crossing the UVCS slit during the CME detection. In particular, we searched for high speed events (> 700 km/s) with a metric type II radio burst detected in the time window from the CME onset to the end of UVCS observations. In order to have the best spatial coverage for events observed at different wavelengths, we searched for a CME also detected in the low corona by the MLSO/Mk4 ground-based coronagraph. The dynamics of the CME was inferred through LASCO coronagraph white-light data (f.o.v. 2.3 – 30 R☉), Fig.1, and MLSO/Mk4 coronagraph data (f.o.v. 1.1 – 2.8 R☉), while the shock dynamics was investigated using UV spectrographic observations of UVCS, Fig. 2, and ground-based radio spectral observations of Culgoora (Australia), Fig. 3, and Holloman (New Mexico). The CME was first detected around 22:30 UT in the O VI 1031.93-1037.62 Å doublet spectral lines by UVCS that at the time of the CME passage was performing a sit-and-stare observation at 1.7 R☉ at a polar angle of 245°. Later, it was also detected by the LASCO instrument at 23:06 UT, propagating beyond the southwest limb of the Sun. The associated type II radio emission was finally detected by ground-based radio spectrographs from about 22:35 to 23:10 UT.

## Data reduction and analysis

From the analysis of the spectral emission of the oxygen doublet O VI 1032 / O VI 1037 Å observed by UVCS, we detected a significant increase of the O VI line intensities and a line broadening of about 30% in a portion of the UVCS slit. These observational features are attributed to plasma compression and ion heating, as expected at the transit of a coronal shock propagating ahead of the front of a fast CME [11]. In general, the observed O VI intensity $I$ is given by the sum of a collisional component $I_{coll} \propto \int_L n_e^2 dl$ and a radiative scattering component $I_{rad} \propto I_{disk} \int_L n_e F_D(\vec{v}) dl$, where $I_{disk}$ is the exciting disk intensity, $n_e$ the electron density and $F_D$ the Doppler Dimming function which depends on the ion outflow velocity, $v$. The observed lines intensities ratio, R=$I_{OVI\ 1032}/I_{OVI\ 1037}$, is expected to range between the value of 4 -in the case of totally radiative contribution- and the value of 2 -in the case of totally collisional contribution- [12]. From the O VI doublet intensity diagnostics, as shown in Fig. 4, we verified that at the shock passage (at ~ 22:40 UT) R tends to the value of 2 due to the rapid acceleration and compression of the plasma. Yet, R still remains > 2 because of the line-of-sight (l.o.s.) contribution of fore/back-ground material, as expected. At the CME transit (at ~ 23:10 UT) R ~ 2 due to the high density of the expanding CME plasma that fills a much greater portion of the l.o.s. In order to estimate the shock speed, $V_{shock}$, from the type II frequency drift observed in the radio dynamic spectra, we inferred the local coronal density profile $n_e(r)$ through inversion of the polarized brightness *(pB)* as derived from the white-light coronagraph observations, thus obtaining a shock speed of ~ 800 Km/s. For the determination of the plasma density we have used the on-line available *pB* calibrated data from MLSO/Mk4 instrument (specifically the one at 22:30 UT of 11.01.2003, just before the CME detection). With the Van de Hulst's inversion method (assuming spherical symmetry, hypothesis generally accepted during solar maximum), we derived a radial density profile in the region of the shock propagation. This profile, that was fitted with a third degree polynomial, was found to be quite similar to a typical Newkirk 1967 profile ($n_e(r) = 4.2 \; 10^4 \; 10^{[4.32/r]}$) that, by the way, is generally the one used in literature in order to determine shock velocities above streamer from radio dynamical spectra. On the basis of the LASCO and MK4 observations, we further estimated the kinematics of the CME front, finding a CME speed of ~750 km/s (i.e. the shock propagates faster by about 6% than the CME itself).

**Results and discussion**

A band-splitting of the emission lines is often observed in the radio dynamic spectra and it is attributed to emission behind and ahead of the shock front [13,14]. From the squared ratio of the frequencies of the two split bands, as observed in the metric radio dynamic spectra, we were able to estimate the compression ratio $X = \frac{f_h^2}{f_l^2} = \frac{n_h}{n_l} = 1.39$. Assuming perpendicular propagation of the shock with respect to the ambient magnetic field, we obtained the Mach magnetosonic number $M_{ms} \equiv \frac{V_{shock}}{V_{ms}}$. In the corona the sound speed is much less than the Alfvén speed $V_A$, so that $V_{ms} \sim V_A$. Consequently, $M_{ms} \cong M_A \equiv \frac{V_{shock}}{V_A}$, and from the Rankine-Hugoniot relationships for $\beta \ll 1$ (that is, the ratio of the plasma pressure to magnetic pressure), yielding $M_A = \sqrt{\frac{X(X+5)}{2(4-X)}} = 1.30$, we estimated $V_A \approx \frac{V_{shock}}{M_A} \approx 610$ km/s. With the above information, by using the electron density obtained from the inversion of the *pB*, we were able to compute the coronal magnetic field strength, decreasing from about 4.4 to 1.5 G in the height range from 1.2 to 1.5 R$_\odot$, as shown in Fig. 5.

## Conclusions

We have analyzed a CME/shock event observed by UVCS/SOHO in the ultraviolet on November 1$^{st}$ 2003. The data analysis, performed also with the aid of radio and visible observations from ground- and space-based instruments, allowed us to obtain estimates of important properties of the coronal plasma crossed by the shock (basically the Alfvén speed $V_A$ = 610 km/s and the magnetic field strength B = 4.4-1.5 G) that were found to be consistent with estimates that were previously inferred with different techniques at comparable heights in the corona.

**Acknowledgements** The authors thanks the organizers of IAGA-III symposium Dr. A. Hady and Dr. L. Damé. Mauna Loa Solar Observatory (MLSO) is operated by the High Altitude Observatory (HAO). SOHO is a project of international cooperation between ESA and NASA. The LASCO CME Catalog is generated and maintained by NASA and Catholic University of America in cooperation with the Naval Research Laboratory.

**Figures**

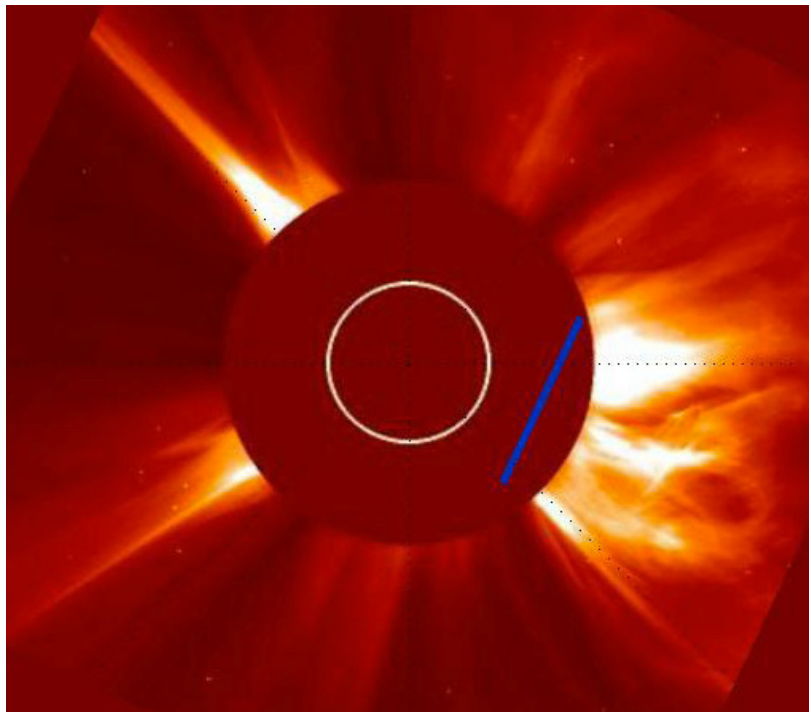

Fig.1: LASCO C2 white light CME image detected on November 1[st] 2003 at 23:54 UT with superposed the UVCS slit position during the observation from 20:07 UT to 01:54 UT in the following day (blue stripe).

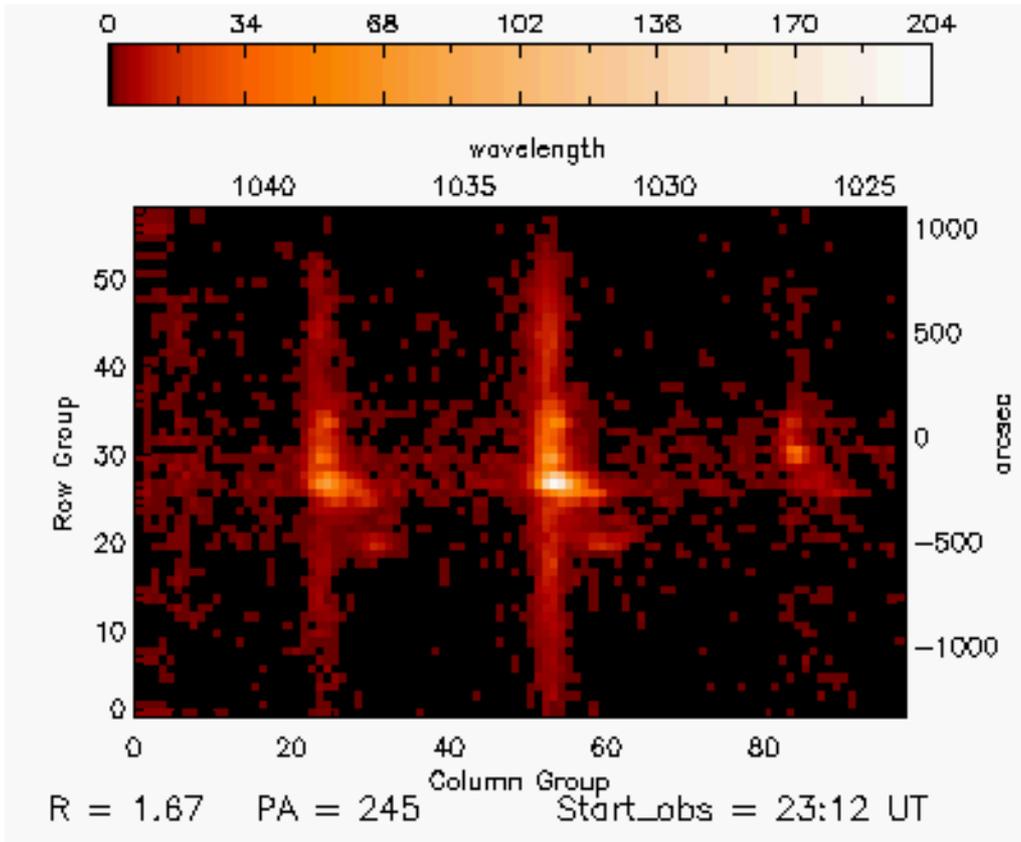

Fig. 2: UVCS spectra of the oxygen doublet O VI 1031.93-1037.62 Å (the two brightest lines) and of the HI Ly$\beta$ 1025.72 Å spectral line.

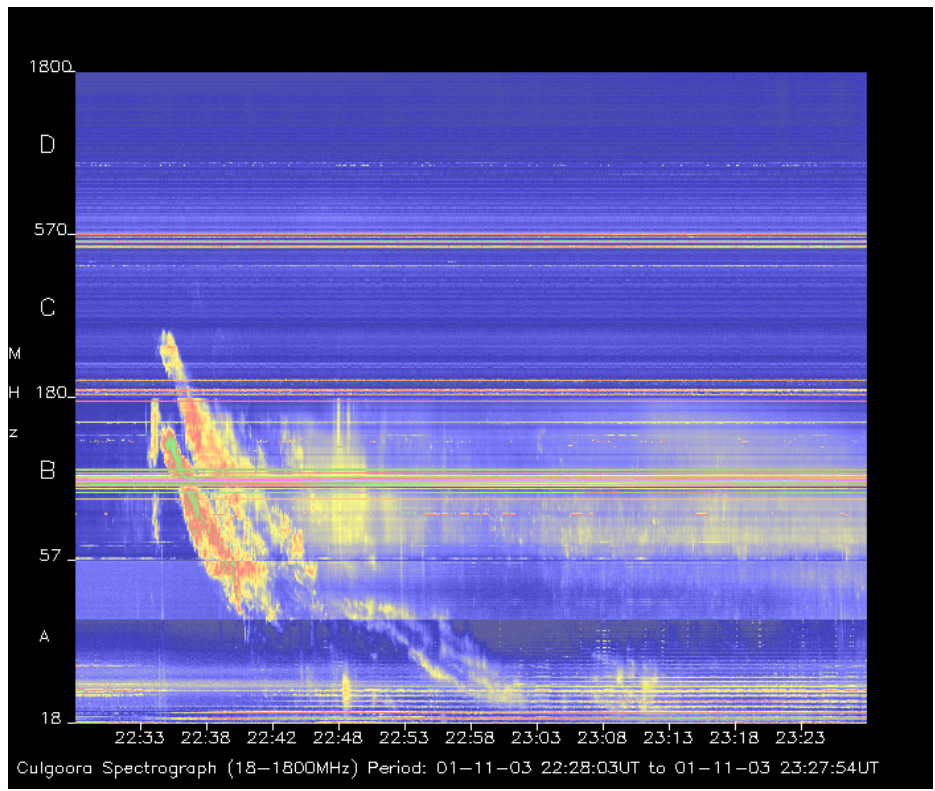

Fig. 3: Dynamic spectrum from Culgoora showing the type II radio burst.

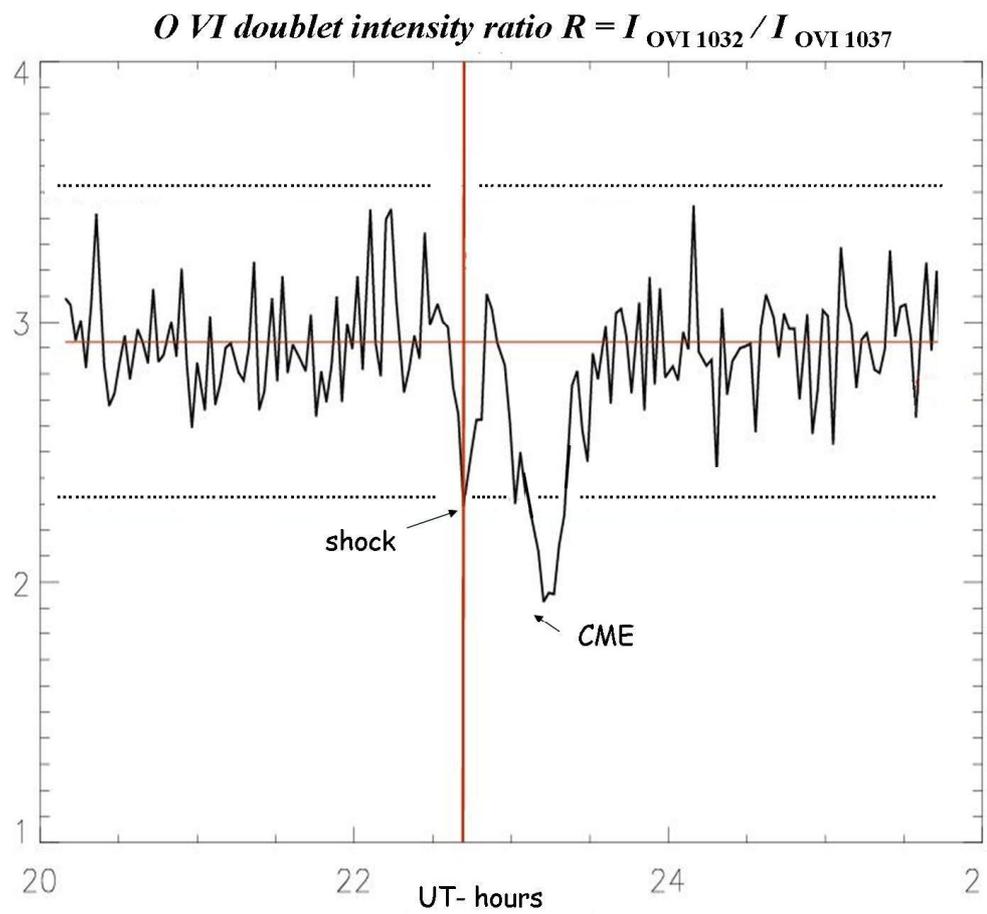

Fig. 4: Observed $I_{OVI\,1032}/I_{OVI\,1037}$ ratio from November 1st (starting at 20:00 UT) and November 2nd, 2003 (ending at 02:00 UT).

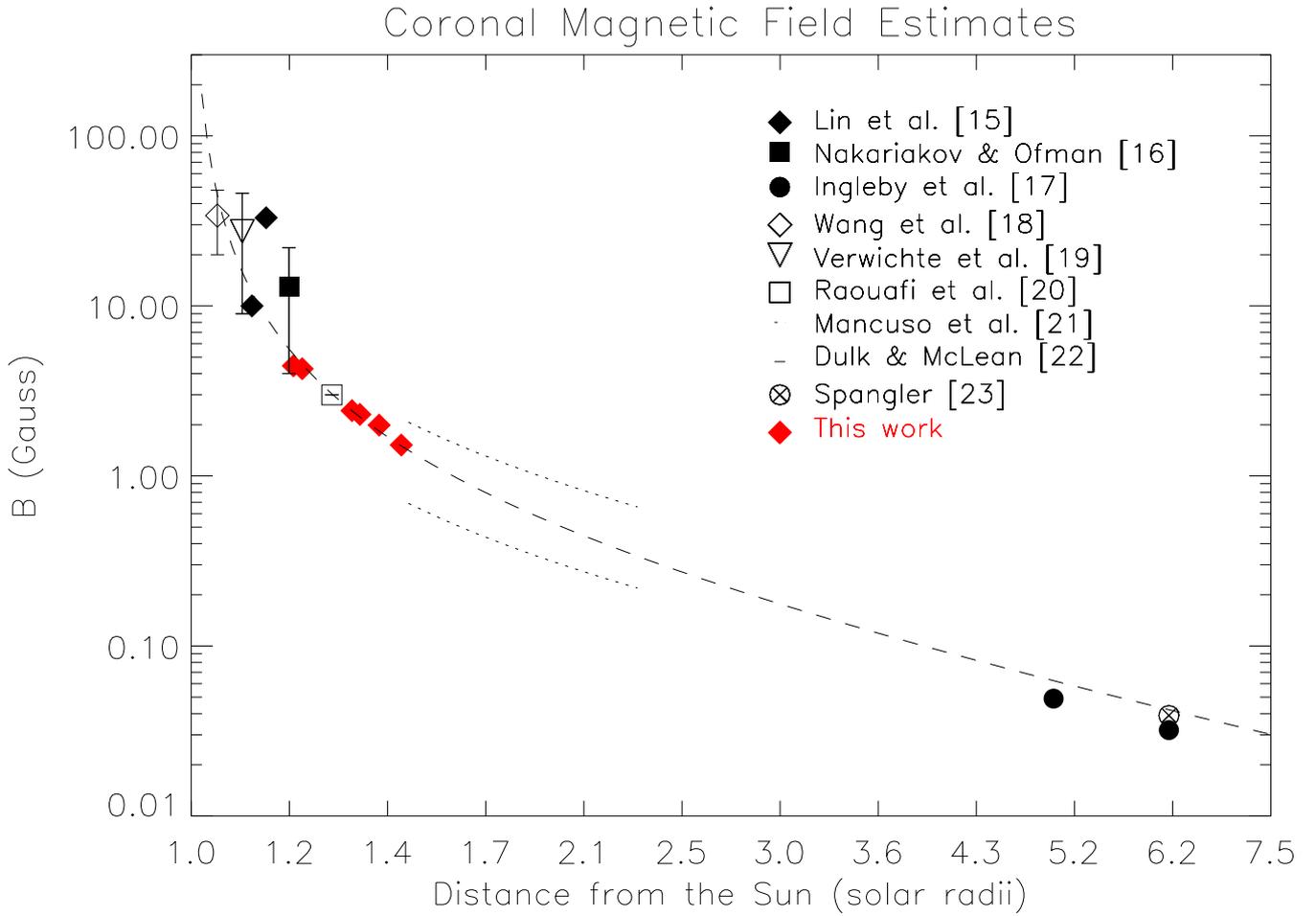

Fig. 5: Estimated values of the coronal magnetic field B strength in the interval 1.2-1.5 $R_\odot$ of the solar corona (plotted in red), compared with some other published estimates (plotted in black) [15-23].